# Enhanced Thermoelectric Properties By Embedding Fe Nanoparticles Into CrN Films For Energy Harvesting Applications

Daria Pankratova[1], Khabib Yusupov[2], Alberto Vomiero[1,3], Sanath Kumar Honnali[2], Robert Boyd[2], Daniele Fournier[4], Sebastian Ekeroth[2], Ulf Helmersson[2], Clio Azina[5], Arnaud le Febvrier[2*]

[1] Department of Engineering Sciences and Mathematics, Luleå University of Technology, 97187 Luleå, Sweden

[2] Department of Physics, Chemistry, and Biology (IFM), Linköping University, Linköping SE-581 83, Sweden

[3] Department of Molecular Sciences and Nanosystems, Ca' Foscari University of Venice, Via Torino 155, 30172 Venezia Mestre, Italy

[4] Sorbonne Université, CNRS, Institut des NanoSciences de Paris, UMR 7588, Paris 75005, France

[5] Materials Chemistry, RWTH Aachen University, Kopernikusstraße 10, D-52074, Aachen, Germany

*Corresponding author: arnaud.le.febvrier@liu.se

# Abstract

Nanostructured materials and nanocomposites have shown great promise for improving the efficiency of thermoelectric materials. Herein, Fe nanoparticles were imbedded into a CrN matrix by combining two physical vapor deposition approaches, namely high-power impulse magnetron sputtering and a nanoparticle gun. The combination of these techniques allowed the formation of nanocomposites in which the Fe nanoparticles remained intact without intermixing with the matrix. The electrical and thermal transport properties of the nanocomposites were investigated and compared to a monolithic CrN film. The measured thermoelectric properties revealed an increase in the Seebeck coefficient, with a decrease of hall carrier concentration and an increase of the electron mobility which could be explained by energy filtering by internal phases created at the NP/matrix interface. The thermal conductivity of the final nanocomposite was reduced from 4.8 W m$^{-1}$K$^{-1}$ to a minimum of 3.0 W m$^{-1}$K$^{-1}$ W. This study shows prospects for the nanocomposite synthesis process using nanoparticles and its use in improving the thermoelectric properties of coatings.

# Keywords:

# Table of content

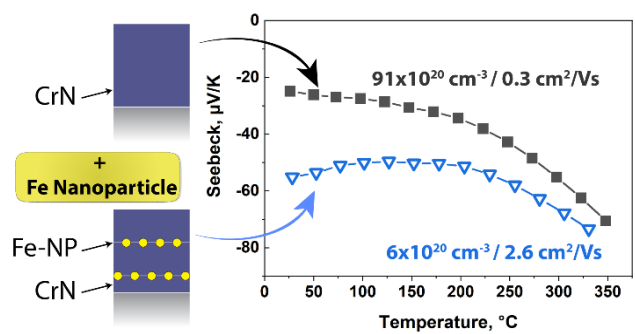

# Introduction

Thermoelectric thin films have shown their potential role in waste heat recovery and microdevice refrigeration [1]. The conversion efficiency of thermoelectric material is often governed by the dimensionless figure of merit, $ZT = (S^2\sigma)T/\kappa$, where S is the Seebeck coefficient, σ is the electrical conductivity, κ is the thermal conductivity, and *T* the temperature [2]. Several strategies exist to improve thermoelectric performance, including the search for new materials or optimization of existing ones by doping, alloying, and/or introducing nanoscale effects [2-5]. Nevertheless, the most prominent challenge for enhanced thermoelectricity is decoupling the three physical parameters (S, σ, and κ) which are strongly interdependent and coupled with carrier concentration [5].

Transition metal nitride (ScN and CrN) systems have shown promise for use as thermoelectric materials [6]. CrN-based materials, in particular, can exhibit both n-type or p-type semiconducting behavior, depending on doping and nitrogen stoichiometry [7-11]. The n-type CrN has a relatively high Seebeck coefficient (60-300 μVK$^{-1}$), high electrical conductivity (0.6-3 mΩ.cm), and medium to low thermal conductivity (4-6 Wm$^{-1}$k$^{-1}$) [6-7, 10]. Due to the presence of magnetic disorder in CrN [12], lower thermal conductivities (~ 2 - 3 W·K$^{-1}$·m$^{-1}$ at room temperature (RT) [6, 11]) were reported compared to other transition metal nitrides with the same crystallographic structure (ScN, VN, TiN: 10 - 22 W·K$^{-1}$·m$^{-1}$ at RT [6, 13-14]).

Several theoretical and experimental studies have suggested new horizons for improving the thermoelectric properties by phonon and carrier engineering using nanoparticle (NP)-containing nanocomposite materials [5, 7, 15-19]. In fact, it was suggested that using soft-magnetic, typically Fe-based, NPs dispersed in a semiconductor matrix would: i) reduce the thermal conductivity by inserting defects which combined with a Kondo-like effect can increase phonons scattering; ii) increase the Seebeck coefficient by electron filtering at the metal/semiconductor interface [20]; and iii) increase the electrical conductivity by modulation doping at the metal/semiconductor interfaces due to the difference in work functions between matrix and NPs [15].

(Nano)composite materials are characterized by (nano)fillers embedded in a matrix which can either be composed of the same phase (e.g., SiC/SiC), or a different one. To avoid intermixing of the fillers and the matrix, phases that are immiscible or that interact poorly are often easier to synthesize. For example, nano-inclusions and nanoparticles have been reported to form during synthesis and were shown to improve the thermoelectric properties of the nanocomposites [7, 9, 18, 21-22]. However, challenges remain in synthesizing (nano)composites in which the matrix and (nano)filler react with each other [23]. To date, there are few reports on the design and synthesis of nanocomposite films containing nano-sized fillers. In fact, these material systems often contain compounds that would normally intermix when processed using conventional synthesis methods with high temperature under more thermionically favorable conditions [24-25]. Such methods can therefore not be used to produce tailor-made nanocomposites.

In the present study, we explore an uncommon synthesis strategy, using physical vapor deposition for growing nanocomposite materials composed of Fe-NP in a CrN thermoelectric matrix. Fe NPs were selected to maximize the thermal fluctuation of magnetic moment as it is the strongest soft-magnetic single element metal, to improve the thermoelectric properties of CrN. nanocomposite coatings consisting of sequentially deposited CrN layers and dispersed Fe-NPs were developed in the same deposition chamber without breaking the vacuum during growth. Using High Power Impulse Magnetron Sputtering (HiPIMS), polycrystalline CrN was deposited at room temperature while the NPs were synthesized using a NP cluster source in a so-called NP gun using a hollow cathode target. To compare the effect of Fe-NPs on the thermoelectric properties of CrN, three coatings were deposited: a CrN film without NPs, and two nanocomposite coatings with different proportions of Fe-NPs. The development of the combined HiPIMS/NP gun approach is shown to be very effective for the synthesis of nanocomposites that cannot be obtained otherwise. Finally, the effect of Fe-NP insertion on the structure, morphology and thermoelectric properties of the coatings was investigated.

# Methods

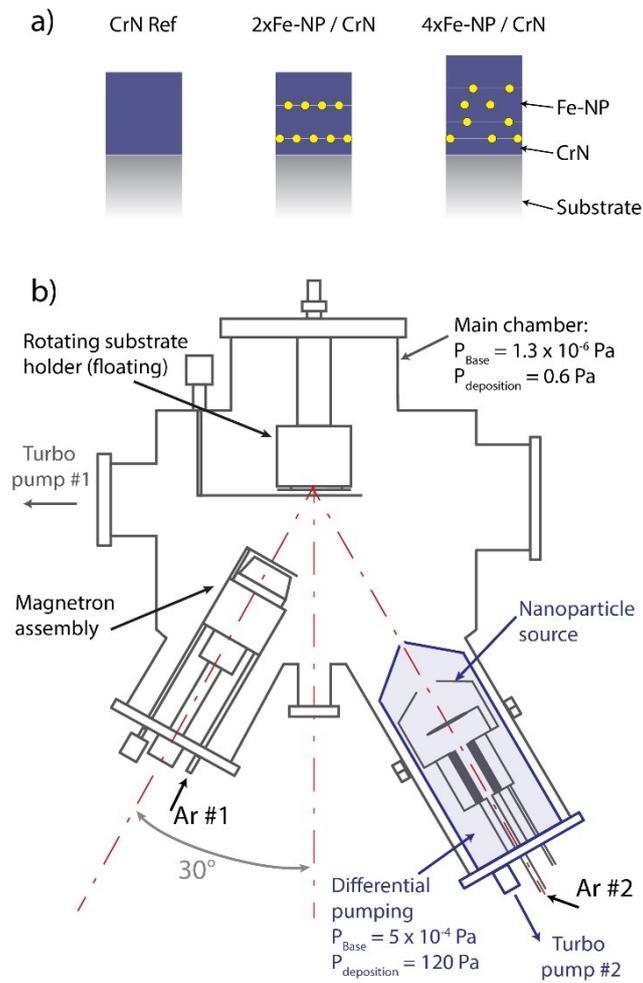

**Figure 1:** Schematic representations of (a) the nanocomposite coatings and (b) the deposition chamber with the magnetron assembly and the NP source in a secondary chamber.

Figure 1 shows a schematic representation of the nanocomposite coating designs and the deposition chamber with the magnetron assembly and NP source. The CrN/Fe-NP coatings consisted of alternating CrN layers and dispersed Fe-NP layers. With a total coating thickness of 80-95 nm, three different types of samples were deposited: i) a reference monolithic CrN film; ii) a nanocomposite coating with two "layers" of Fe-NP named 2xFe-NP/CrN film; and iii) a nanocomposite coating with four layers of Fe-NP named 4xFe-NP/CrN film (Figure 1a). The thickness of each CrN layer was controlled by tuning the deposition time where the bottom layer was kept constant at 15 nm and the top layer at 30 nm. The layers of CrN between the Fe NPs were fixed at 30 nm and 15 nm for the 2xFe-NP/CrN film and 4xFe-NP/CrN film, respectively. The nanocomposites were deposited without intentional heating to minimize diffusion and solubility of Fe into CrN. The lack of temperature during growth was overcome by using HiPIMS and its highly energetic ions sputtered particle which allows to produce high crystalline epitaxial film even without intentional heating [26].

The CrN thin film was deposited by HiPIMS in an ultra-high vacuum chamber with a base pressure of $1.3\times10^{-6}$ Pa using a 50 mm diameter Cr target (Kurt J. Lesker / 99.95%). More details on the deposition chamber can be found elsewhere [27]. The chamber was equipped with one magnetron assembly and a NP source (Figure 1b). The magnetron assembly, situated at 135 mm from the substrate, was powered by a HiPIMS unit (HiPSTER 1, Ionautics AB) at an average target power of 50 W, an average peak current of 8.5 A, and fixed voltage, pulse length, pulse frequency of 500 V, 30 µs, 700 Hz, respectively. The deposition conditions of the CrN layer mentioned above were chosen to obtain pure NaCl-B1 phase at room temperature.

The NP source consists of a "secondary cylindrical chamber" with a 1 mm orifice directed toward the substrate and placed at 250 mm from the substrate. The "secondary cylindrical chamber" with an approximate volume of 0.00225 m$^3$ (dimension: 26 cm × 10.5 cm ⌀) had a differential pumping with a base pressure of $5\times10^{-4}$ Pa (Figure 1b) in which a hollow cathode sputtering system was placed to produce NPs. Note here that the walls of the secondary chamber were electrically floating to avoid any disruption in the NP synthesis and deposition. The NP source (HiPNano, Ionautics AB) was equipped with a Fe hollow cathode of $\varnothing_{outer}$ 9 mm/$\varnothing_{inner}$ 5 mm × 54 mm (Good Fellow, 99.5 % purity), and the anode ring was fixed 10 mm away from the hollow cathode. The NP source was powered by a HiPIMS unit (HiPSTER 1, Ionautics AB) at an average power of 20 W, pulse length of 30 µs, pulse frequency of 300 Hz and average peak current of 13 A and voltage of 485 V. The conditions of deposition and anode position in the NP source were selected to obtain Fe-NP of 6-13 nm in diameter. The total NP areal density per micrometer square was kept constant at 800 NP/µm$^2$ in both nanocomposites. The NP areal density in one layer (duration of deposition for one layer) was changed from 400 NP/µm$^2$ (10 min) for 2xFe-NP/CrN film and 200 NP/µm$^2$ (5 min) for 4xFe-NP/CrN film. The areal density and size estimation were performed by scanning electron microscopy.

A gas mixture consisting of 16 sccm of Ar and 5 sccm of N$_2$ was inserted in the main chamber while pure Ar gas was injected through the hollow cathode with a 28 sccm flow. The differential pumping in the NP source was throttled to reach a pressure of 130 Pa in the NP gun, and a pressure of 0.6 Pa in the main chamber.

One-sided polished c-plane sapphire substrates with dimensions of 10×10×0.5 mm$^3$ were used. The substrates were cleaned with acetone and ethanol in ultrasonic baths for 10 min and were blow-dried using N$_2$ before being mounted on the substrate holder which was kept at room temperature and under constant rotation during depositions.

The crystal structure of the films was analyzed using a PANAnalyical X'pert powder diffractometer in a Bragg Brentano configuration using a Cu-Kα radiation wavelength of 1.5406 Å (45 kV and 40 mA). A nickel filter was used to filter the K$_\beta$ signal visible on the substrate peak. The measurement scan was recorded with constant rotation of the sample using a X'celerator detector in 1D scanning line mode: 10-80° 2θ range, 0.0084° step size, equivalent of 19.7 sec/step.

The elemental composition of the films was performed by ion beam analysis using the Pelletron Tandem accelerator (5MV NEC-5SDH-2) at the Tandem Laboratory, Uppsala University, Sweden [Ref]. Time-of-flight elastic recoil detection analysis (ToF- ERDA) measurements were conducted using 36 MeV iodine ion ($^{127}$I$^{8+}$) beam. The incident beam angle to the target surface normal was 67.5°, while the ToF-telescope and the gas ionisation detector were placed at 45° relative to the incident beam direction. The depth profile of the elemental composition was acquired from the ToF-ERDA time energy coincidence spectra using the *POTKU* 2.0 software [28].

A scanning electron microscope (ZEISS Gemini SEM 650) operated with an acceleration voltage of 2 kV and an in-lens detector was used to observe surface morphologies. Thin lamellae were prepared in a FEI Helios NanoLab dual-beam focused ion beam (FIB) microscope, using Ga$^+$ ions accelerated at 30 kV, using the standard lift out and thinning procedure. All scanning transmission electron microscopy (STEM) analyses were performed on a FEI Tecnai G2 microscope operated at 200 kV. STEM imaging was conducted in high angle annular dark field (HAADF) and was combined with energy dispersive X-ray spectroscopy (EDS).

The transport properties, *i.e.,* in-plane electrical conductivity and in-plane Seebeck coefficient were measured by the 4-point probe-measurement method in the 25-325 °C temperature range using a NETZSCH SBA 458 nemesis. Before the measurement, the chamber was filled with Ar using the standard procedure, i.e., the vacuum was applied to the chamber via pumping and filled with Ar. This procedure was applied three times to ensure the neutral environment of the chamber. To further evade the risk of pressure accumulation while heating, the chamber was purged with Ar gas at the rate of 50 ml/min. During the measurements, two microheaters generated the temperature gradient in both sample directions (one sample side being heated and cooled down, and after that, the same applies to another sample side).

A second electrical setup was used to evaluate the resistivity, drift mobility and Hall carrier concentration at room temperature. This homemade setup consists of a Van der Pauw configuration setup adapted for Hall measurement. A Keithley 2400 SourceMeter® and a Keithley 2100 Multimeter® were used to track the current values and measure voltages, respectively. A permanent magnet of 0.5 T was used leading to a magnetic field of 0.485 T at the sample position. The Hall measurement was conducted with a constant current of 40 mA where a Labview program was used to select the different measurement configurations necessary for the Hall measurement using the Van der Pauw configuration.

The thermal conductivity of the films was obtained at room temperature using modulated thermoreflectance microscopy (MTRM). In this setup, a pump beam (532 nm) delivered by a Cobolt MLD laser, intensity-modulated by an acousto-optical modulator at a frequency of $f$, is focused on the surface of the sample with an objective lens (N.A. = 0.5). The variation of surface reflectivity was monitored by a probe laser (638 nm Oxxius laser). A photodiode and a lock-in amplifier record the AC reflectivity component at a frequency of 100 kHz. The specification of the setup is the spatial measurement around the pump beam. The measurement of the reflectivity of the probe on the surface is performed along the x-axis from -10 μm to + 10 μm around the pump beam area. Finally, the amplitude and phase data were fitted according to a standard Fourier diffusion law to extract the thermal conductivity of the CrN and nanocomposite films [29-32].

The different characterization techniques were performed on the same samples, and in the following order: XRD, SEM, in-plane electrical resistivity and in-plane Seebeck coefficient from RT to 350 °C, modulated thermoreflectance microscopy (MTRM) and finally EDS-STEM. It is important to note that the thermal conductivity and TEM were performed after the electrical characterization measurements which included heating of the samples to up to 350 °C.

# Results and discussions

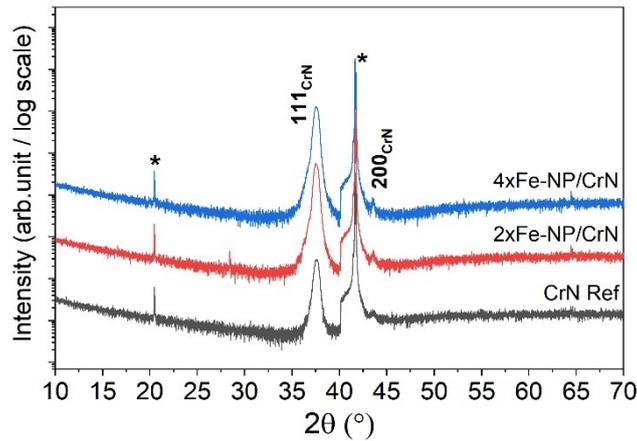

**Figure 2:** θ-2θ XRD patterns of the monolithic reference CrN film and the two nanocomposite coatings with 2 and 4 layers of Fe-NP.

Figure 2 presents the Bragg Brentano θ-2θ XRD patterns collected for the reference CrN film and the two nanocomposite coatings. The peaks from the c-plane sapphire substrate were observed at 2θ of 20.5° and 41.6° corresponding to the 0003 and 0006 reflections of $Al_2O_3$, respectively. A sharp cut-off feature at around 40° can be observed on the substrate peak and is due to the use of Nickel filter in front the detector. The two other XRD peaks observed belong to CrN 111 and 200 with 2θ = 37.6° and 2θ = 43.5°, respectively. CrN showed similar features in all films with a preferential orientation along the [111] direction with a minor quantity of grains oriented along the [200] directions. This preferential orientation of NaCl-B1 CrN on c-plane sapphire is not surprising and has been observed previously [7]. The cell parameter of CrN was estimated at 4.15 Å for all coatings and was consistent with the experimental and theoretical values reported for CrN (4.13 Å) [33-34]. For the nanocomposite, only the CrN phase was observed as no peaks from Fe-bcc were detected. That is reasonable due to the small volume and size of the Fe NPs. No significant differences were noticeable between the two nanocomposite films except for the relative intensity of the 111 diffraction peak compared to the 200 diffraction peak. Those differences, observable on a logscale, remain low and may originate from variations in the proportion of preferentially oriented grains in the film. Despite the low temperature of deposition and the relatively low thickness of the film, clear diffraction peaks were detected revealing the presence of crystalline of CrN phase in the film.

**Table 1:** Composition of the film measured by ToF- ERDA.

| Sample | Cr (at. %) ± 0.5 | N (at. %) ± 0.5 | O (at. %) ± 0.5 | $CrN_{1-\delta}$ with $\delta \pm 0.01$ |
|---|---|---|---|---|
| CrN ref | 54.8 | 42.2 | 3.0 | $CrN_{0.77}$ |
| 2xFe-NP/CrN | 54.2 | 41.8 | 4.1 | $CrN_{0.77}$ |
| 4xFe-NP/CrN | 54.7 | 41.9 | 3.4 | $CrN_{0.77}$ |

Table 1 displays the composition of the films measured by ToF-ERDA. All three samples have the same composition within the error bar of measurement and are understoichiometric in nitrogen ($CrN_{0.77 \pm 0.01}$). No Fe was detected as the content is below the detection limit of ToF-ERDA. Considering a concentration of Fe-NP of 800 NP/µm$^2$ with a density of 7.87 g/cm$^3$ (bcc-Fe ICDD 00-006-0696) and a density of CrN of 5.89 g/cm$^3$ (Na-Cl-CrN ICDD 01-074-8390), the percentage of iron was estimated to be 0.2 at. % overall, and 0.4 at. % to 0.8 at. % within the Fe-layer (average of 10 nm thickness) for 4xFe-NP/CrN and 2xFe-NP/CrN films, respectively. A 3-4 at. % oxygen contamination was detected for all the films which was most probably incorporated during the deposition.

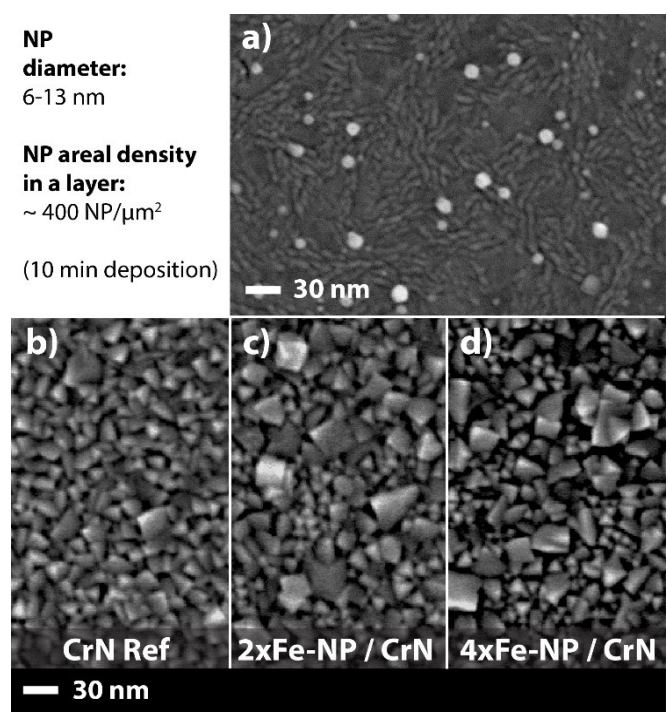

**Figure 3:** Top surface morphology of a) test sample for NP for areal density/size estimation. The NPs were deposited on a Cr layer for a duration of 10 min; b) the monolithic CrN films and the two nanocomposite coatings with c) 2 and d) 4 layers of Fe-NP.

Figure 3a) shows the surface morphology of a test sample produced to determine the NP areal density and estimate the sizes of the deposited NPs. The NPs had sizes ranging from 6 and 13 nm with an areal density of 400 NP/µm$^2$ for 10 min deposition duration, approximately. The reference film, and the two nanocomposite coatings (2xFe-NP/CrN and 4xFe-NP/CrN) are also shown in Figure 3b), c), and d).. The reference CrN films exhibited the typical morphology of cubic NaCl B1 structure materials deposited on sapphire [35-36], which consists of pyramid-shaped grains (10-15 nm in size) consistent with the XRD results and the (111) oriented film. Locally few different grain orientations were visible with square-shaped or elongated grains corresponding to the secondary (200) orientation. The nanocomposite coatings had similar characteristics as the reference sample but with increased numbers of randomly oriented grains. The presence of more square-shaped grains and/or elongated grains is consistent with the XRD observations. The presence of Fe NPs could be responsible for disturbing, to some extent, the growth of the CrN coating as seen by the more pronounced randomly oriented character of the CrN surface morphology compared to the reference sample. Nevertheless, their presence was not enough to drastically change the morphology such as yielding porous films.

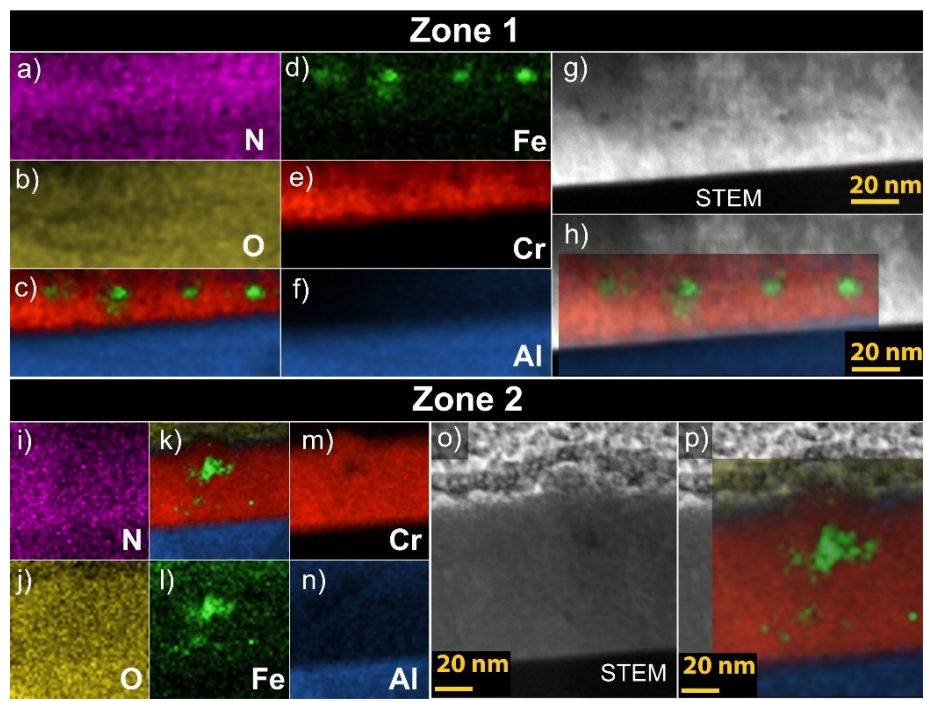

**Figure 4:** STEM analyses of the 4xFe-NP/CrN film in two zones of interest. Zone 1 is for NPs at the bottom part of the film while zone 2 is for NPs at the top part of the film. (a-f) and (i-n) EDS maps of the region of interest with the map corresponding to N, O, Fe, Cr, and Al, respectively. (d, o) STEM images of the zone of interest, and (e, p) superposition of the EDS map on the STEM image.

Figure 4 shows the STEM images and EDS maps of two areas of interest from the 4xFe-NP/CrN film (noted zone 1 and zone 2). The NPs spatial distribution and small size means that when observing the film's cross section there will always be overlap between the particles and the matrix, even within a thin TEM specimen. Individual particles are then almost impossible to see using traditional transmission electron microscopy (TEM) dark field TEM and selected area electron diffraction. However, 6-9 nm sized features were observed in STEM images (figure 4 zone 1). While Zone 1 corresponds to the first NP layer at the bottom part of the film, Zone 2 represents an area at the top part of the film where several NPs seemed to locally form a cluster. Subsequent EDS-STEM analyses show that these features have a high Fe content, indicating the presence of intact nanoparticles with no clear differences of N or O along the films and compared to the matrix. Note that EDS-STEM is not sensitive to light elements and distinctions between C, N and O cannot be resolved making the interpretation challenging. Nevertheless, the features correspond to nanoparticles of Fe, estimated to range between 6 and 9 nm, which is consistent with the observations of the Fe-NP morphology made by SEM (Figure 3).

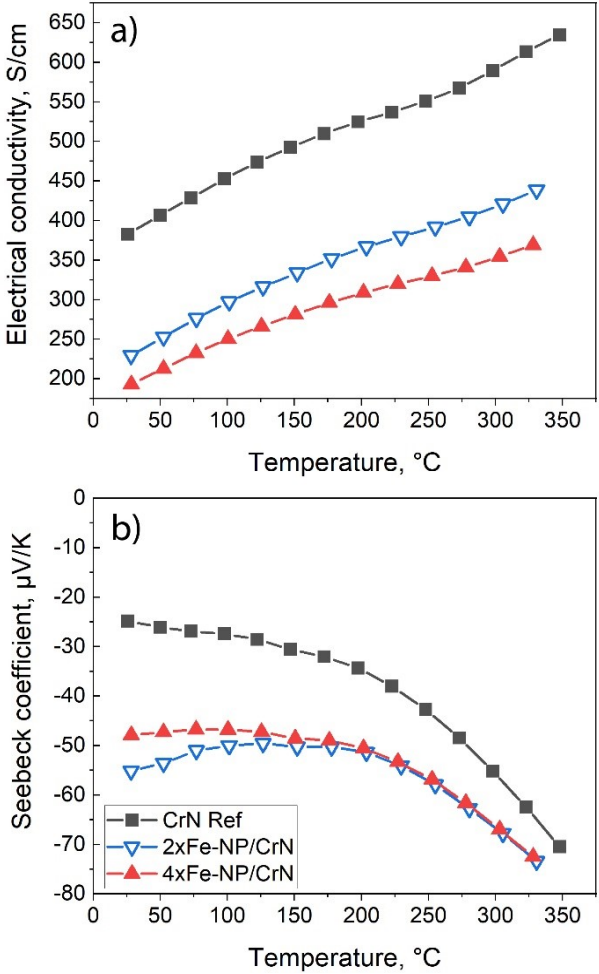

**Figure 5:** Temperature-dependent thermoelectric properties: a) electrical conductivity and b) Seebeck coefficient measured on the CrN reference coating (filled squares) and the two nanocomposite coatings (2xFe-NP/CrN, empty triangles and 4xFe-NP/CrN, filled triangles).

Figure 5 displays the thermoelectric properties of the three coatings. In the case of the CrN reference sample, both absolute Seebeck coefficient and electrical conductivity values increase with temperature from -25 to -70 µV/K and from 375 to 625 S/cm, respectively, which is a typical behavior of semiconductors. The low Seebeck coefficient values compared to other studies could be caused by the low thickness of the film and small CrN grains (<100 nm) which constitute the matrix [7, 37-39]. In addition, the N-understoichiometric composition of the film could also influence the Seebeck coefficient of the reference sample where an understoichiometry reduces the Seebeck values [27]. The reported room temperature electrical resistivity values exist in a wide range for CrN layers, ranging over 2.5 orders of magnitude from $1.7\times10^{-3}$ to $3.5\times10^{-1}$ $\Omega$cm [38-44]. Most studies suggest a strong correlation of the resistivity of the coatings and the crystal quality/morphology of the coating, where higher quality/larger grain reduce the electrical resistivity and increases the Seebeck coefficient [9, 11, 37, 45]. The CrN sample produced in this study was characterized by an electrical resistivity of 2.54 m$\Omega$cm which falls within the lower range of resistivities mentioned above.

Compared to the CrN film, the nanocomposites exhibited higher absolute Seebeck values of 50-60 $\mu$V/K and lower electrical conductivities of 200-230 S/cm at room temperature, indicating superior performance compared to the monolithic CrN film. In addition, their temperature dependence resembles that of the reference CrN, as all samples exhibited a semiconducting behavior with an increase of the electrical conductivity with temperature.

**Table 2:** Room temperature thermoelectric characteristics: electrical resistivity, Seebeck coefficient, charge carrier concentration and mobility, and total, electronic and lattice thermal conductivities.

| Sample | Thickness (nm) | Resistivity (m$\Omega$ cm) ± 0.05 | S ($\mu$V K$^{-1}$) ± 2 | S$^2\sigma$ ($\mu$W m$^{-1}$ K$^{-2}$) | $\mu_{Hall}$ (cm$^2$ V$^{-1}$ s$^{-1}$) ± 0.1 | n ($\times 10^{20}$ cm$^{-3}$) ± 0.1 | Thermal conductivity (W m$^{-1}$ K$^{-1}$) | | |
|---|---|---|---|---|---|---|---|---|---|
| | | | | | | | $\kappa_{Total}$ | $\kappa_e$ | $\kappa_l$ |
| CrN ref | 80 | 2.54 | - 26 | 26 | 0.3 | 92.1 | 4.80 | 0.29 | 4.51 |
| 2xFe-NP/CrN | 85 | 3.28 | - 55 | 92 | 2.6 | 7.3 | 3.60 | 0.22 | 3.44 |
| 4xFe-NP/CrN | 90 | 4.26 | - 49 | 56 | 2.2 | 6.7 | 3.00 | 0.17 | 2.83 |

Table 2 lists the electrical and thermal characteristics of the film such as carrier concentration, drift mobility, and thermal conductivity measured at room temperature. The reference sample exhibited a Hall carrier concentration of $\sim$92 $\times$ 10$^{20}$ cm$^{-3}$ and a Hall mobility of 0.3 cm$^2$V$^{-1}$s$^{-1}$. Compared to values reported in the literature (0.2 to 7$\times$ 10$^{20}$ cm$^{-3}$), the measured values of the carrier concentration are rather high while the hall mobility falls within the reported values of 0.1 to 3 cm$^2$V$^{-1}$s$^{-1}$ [7, 9, 38-39]. For both nanocomposites, the Hall carrier concentration dropped by one order of magnitude when the Fe-NPs were embedded into CrN and their electron drift mobilities were one order of magnitude higher than the reference sample. While these values differ from the monolithic sample, they fall in the ranges reported in the literature.

The effect of nano-inclusions on the electrical conductivity has been observed previously on similar material systems [11]. For example, nano-inclusions of metal formation observed in heavy metal (Mo or W) alloyed CrN yielded an increase of the electrical resistivity from 2 m$\Omega$cm (500 S.cm) to 12 m$\Omega$cm (83 S.cm) [11]. That deterioration of the electrical behavior was attributed to the increase of electron scattering at interfaces with the presence of nano-inclusions of metal in the thermoelectric material.

In the present study, the effect of NPs on the electrical behavior and on the Seebeck coefficient cannot be entirely described. Their insertion led to an increase of the electron mobility and of the Seebeck coefficient, and simultaneously reduced the measured Hall carrier concentrations. A decrease of the Seebeck coefficient together with an increase of the electrical conductivity is not surprising and is generally attributed to ionized impurities when the carrier concentration increases. Nevertheless, as compositions of the three films were the same ($CrN_{0.77}$), it is unlikely that ionized impurities lead to the different electrical behaviors of the samples. Moreover, between the two nanocomposites, the maximum Seebeck was observed for the film exhibiting the largest Hall electron mobility and carrier concentration, which was the film containing less Fe-NPs.

The increase of the electron mobility together with the decrease of the Hall carrier concentration could result from energy filtering by internal phases created at the NP/matrix interface. Electron energy filtering can be triggered by energy barriers in non-uniform materials with doping modulation, heterostructures, and composite materials. Enhancement of $S$ and power factor ($S^2\sigma$) have been observed in such structures and was attributed to the presence of energy barriers [46-49]. In such cases, the low-energy electrons are scattered effectively, while high energy electrons travel through NPs resulting in an increase of the electron mobility, a reduction of the Hall carrier concentration and, hence, the increase of the Seebeck coefficient [2, 46, 50].

The measured total thermal conductivity was reduced from 4.8 W m$^{-1}$K$^{-1}$ for a monolithic CrN to a minimum of 3.0 W m$^{-1}$K$^{-1}$ for the 4xFe-NP/CrN film. The lattice thermal conductivity $\kappa_l$ was calculated from the difference between the total thermal conductivity $\kappa_{Total}$, and the electronic part of the thermal conductivity $\kappa_e$, which was deduced according to the Wiedemann–Franz law, $\kappa_e = LT\sigma$, where L is the Lorenz number (2.44 × 10$^{-8}$ W$\Omega$K$^{-2}$) and T is the temperature. In the present case, both electronic and lattice thermal conductivities decrease to contribute to an overall reduction of the total thermal conductivity. It is worth noting that the lattice thermal conductivity represented the major contribution (90%) of the total thermal conductivity reduction and was reduced by 25 % for the 2xFe-NP/CrN film and by 40 % for the 4xFe-NP/CrN film. The decreased thermal conductivity was attributed to the local interfaces creating phonon scattering centers between the CrN matrix and Fe-NPs. Using NPs to decrease the thermal conductivity of thermoelectric materials has been reported in the literature and similar effects, as the reduction observed in this study, were discussed [18, 51-53].

Despite the lower Seebeck values measured for the nanocomposites compared to values reported in the literature, the nanocomposites outperformed the reference CrN film. This suggests that the addition of Fe-NPs could be beneficial for improving the thermoelectric properties of transition metal films. Further investigations are therefore required to identify which nanocomposite design may yield improved thermoelectric performance.

# Conclusion

A novel approach was employed to produce nanocomposites composed of Fe-NPs dispersed in a CrN matrix. The novel approach involved the use of a NP gun and simultaneous HiPIMS deposition. The presence of Fe-NPs in the CrN matrix was confirmed by STEM and EDS mapping where the Fe NPs remained intact without noticeable reaction with the matrix, which validate the use of our methodology for synthesizing nanocomposites of miscible material systems. The thermoelectric properties of the nanocomposites were assessed and compared to a monolithic CrN reference film. The addition of Fe NPs had an impact on the electrical properties of the films by lowering the Hall carrier concentration, increasing the electron mobility, and yielding an increase of the Seebeck coefficient, while decreasing the electrical and thermal conductivities. The changes in electrical behavior have been attributed to the decrease of the Hall carrier concentration accompanied by an increase of electron mobility, while the decrease of the thermal conductivity with increasing number of Fe-NPs was linked to increased phonon scattering at the newly produced Fe/CrN interfaces.

This study revealed the potential of the nanocomposite strategy using magnetic NPs which could lead to significant improvements in the performance of thermoelectric materials with increased Seebeck coefficient and decreased thermal conductivity.

# Acknowledgements

The authors acknowledge support from the Swedish Government Strategic Research Area in Materials Science on Functional Materials at Linköping University (Faculty Grant SFO-Mat-LiU No. 2009 00971), and the Swedish Research Council (VR) under project grant 2021-03826. A.V. acknowledges the Knut och Alice Wallenberg Foundation (grant number KAW 2016.0346) and the Kempe Foundation for financial support. Daniel Primetzhofer and Mauricio Sortica from Uppsala University are acknowledged for Accelerator operation supported by Swedish Research Council VR-RFI (No. 2019-00191) and the Swedish Foundation for Strategic Research (No. RIF14-0053).